\documentclass[utf8]{frontiersSCNS} 

\setcitestyle{authoryear,open={((},close={))}} 

\usepackage{url,hyperref,lineno,microtype,subcaption}
\usepackage[onehalfspacing]{setspace}

\usepackage{textcomp}
\usepackage{gensymb}
\usepackage{natbib}
\usepackage{graphicx,caption}
\usepackage{float}
\usepackage{comment}

\def\firstAuthorLast{Feldotto {et~al.}, 2022} 
\def\Authors{Benedikt Feldotto\,$^{1,*}$, Cristian Soare\,$^{1}$, Alois Knoll\,$^{1}$, Piyanee Sriya\,$^{2}$, Sarah Astill\,$^{2}$, Marc de Kamps\,$^{3,4,5}$ and Samit Chakrabarty\,$^{2}$}
\def\Address{
  $^{1}$Robotics, Artificial Intelligence and Real-Time Systems, Technical University of Munich, Munich, Germany
  $^{2}$School of Biomedical Sciences, Faculty of Biological Sciences, University of Leeds, Leeds, United Kingdom
  $^{3}$School of Computing, University of Leeds, Leeds, United Kingdom
  $^{4}$Leeds Institute for Data Analytics, Leeds, United Kingdom
  $^{5}$The Alan Turing Institute, London, United Kingdom
}
\def\corrAuthor{Corresponding Author}

\def\corrAuthor{Benedikt Feldotto}
\def\corrEmail{feldotto@in.tum.de}

\begin{document}
\onecolumn
\firstpage{1}

\title[Evaluating Muscle Synergies]{Evaluating Muscle Synergies with EMG Data and Physics Simulation in the Neurorobotics Platform}

\author[\firstAuthorLast ]{\Authors} 
\address{} 
\correspondance{} 

\extraAuth{}

\maketitle

\begin{abstract}

Although we can measure muscle activity and analyze their activation patterns, we understand 
little about how individual muscles affect the joint torque generated. It is known that they are controlled by circuits in the spinal cord, a system much less well understood than the cortex. Knowing the contribution of the muscles towards a joint torque would improve our understanding of human limb control.
We present a novel framework to examine the control of biomechanics using physics simulations informed by electromyography (EMG) data. 
These signals drive a virtual musculoskeletal model in the Neurorobotics Platform (NRP), which we then use to evaluate resulting joint torques. 
We use our framework to analyze raw EMG data collected during an isometric knee extension study to identify synergies that drive a musculoskeletal lower limb model. The resulting knee torques are used as a reference for genetic algorithms (GA) to generate new simulated activation patterns.
On the platform the GA finds solutions that generate torques matching those observed. Possible solutions include synergies that are similar to those extracted from the human study. In addition, the GA finds activation patterns that are different from the the biological ones while still producing the same knee torque.
The NRP forms a highly modular integrated simulation platform allowing these in silico experiments. We argue that our framework allows for research of the neurobiomechanical control of muscles during tasks, which would otherwise not be possible.

\end{abstract}

\section{Introduction}
We take our freedom to move around at will for granted. Only when this freedom is impaired do we start to realize how much so, and only when we try to alleviate the impairment with the aid of technology do we start to fully appreciate how complex the pathway from intended motion to execution actually is.

Much of our normal walking is unconscious, driven by rhythmic neural activation generated in the spinal cord by the central pattern generators. It has long been obvious that the spinal cord has considerable autonomy in calculating appropriate responses to proprioceptive input and that cortical intervention would be too slow. Examples are the changing load of a coffee cup being filled, the precise drawing of a figure automatically adapting to the coarseness and resistance of the canvas, or the fastening of a tiny screw in a delicate machine. In these cases, cortex generated intent passes down by the descending pathways but the generation of precise motor output depends substantially on processing in the spinal cord, integrating and balancing cortical input and proprioceptive feedback.

How this integration is done is still unclear. The neural circuitry of the spinal cord has not been mapped functionally at the same level of detail as that of the cortex. This is because much of the knowledge gained through invasive techniques, e.g. multi-electrode arrays, cannot be performed in humans, except in unusual circumstances. The generalization of ideas that were developed in models of four-legged creatures to human performance is clearly problematic.

A non-invasive technique is electromyography (EMG), where relatively small wearable sensors are used to record muscle activity. As we will see, together with neural modelling, such measurements can provide insight in local processing in the spinal cord. We have experimental evidence that demonstrates so-called \emph{synergies}, the co-activation of several muscle groups to produce a desired motor output. Recently, we have demonstrated that proprioceptive feedback plays a role in the generation of such synergies \citep{york2021effect}, contrary to what has been reported in the literature. Using non-negative matrix factorization in the data analysis, we can identify these synergies. The model produces the same synergy trends as observed in the data, driven by changes in the afferent input. Neural modelling helps interpret these data. To match the activation patterns from each knee angle and position of the study, the model predicts the need for three distinct inputs: two to control a non-linear bias towards the extensors against the flexors, and a further input to control additional inhibition of the rectus femoris. Simulations suggests that proprioception may be involved in modulating muscle synergies encoded in cortical or spinal neural circuits.

What this work does not show and cannot show is how these synergies are involved in the generation of force vectors. Our data shows, strikingly, that individuals can use different synergies to produce the same force vectors. This is a very important observation as understanding this probably helps in understanding the considerable amount of redundancy that is available in the musculo-skeletal system: there is more than one way to produce a given motor output. 

Importantly, even in animal studies there is no access to muscle forces. To understand how force is generated once the activation pattern has been established, simulation is the only available option. The Neurorobotics Platform (NRP) \citep{AloisKnoll.2016} \citep{Albanese.2018} \citep{Hinkel.2016} developed within the Human Brain Project \citep{Amunts.2016} \citep{Markram.2011} is a natural simulation environment for gaining an understanding of force vector output by neural signals, as we can model the entire system: neural activation, resulting muscle tensions and resulting force output, in this experiment the resultant force on a knee torque.

In this paper, we extend the feature rich framework of the NRP for EMG data evaluation and demonstrate that it is ideally suited to investigate force generation in a musculo-skeletal model. Moreover, since it is an entirely simulated environment, this allows \emph{in silico} experimentation. We applied machine learning and signal processing to reanalyze the data trials, determine features and recover the synergies in the signal. We feed the muscle activation into the physics simulation and quantify resulting knee torques. Finally, we use a Genetic Algorithm (GA) to generate alternative control signals that produce the same torque output as in the study. We find that there are various synergies possible. Some of the synergies found in simulation match those identified in the experimental data.  This suggests that a simulated environment can be used to infer potential implementations for control strategies.

\section{Neurorobotics Platform}
The Neurorobotics Platform (NRP)  is developed within the European Human Brain Project  and is integrated into the EBRAINS research infrastructure for brain simulation on supercomputing infrastructure. 
The NRP uses the robot simulator Gazebo \citep{Koenig.2004} for rigid body computations and has been extended with the muscle simulation framework from OpenSim \citep{Seth.2018} \citep{Delp.2007} to simulate musculoskeletal models interacting with virtual environments. This unique integration enables simulation of classical motor-driven robots and biologically derived musculoskeletal mammalian bodies in the same experiment, in our case a musculoskeletal lower limb model in a virtual test environment.
As part of the NRP the Closed Loop Engine is implemented to synchronize spiking neural network simulations in NEST \citep{Gewaltig:NEST} with the physics simulation in predefined simulation steps. Transfer Functions specify the bidirectional exchange of sensory data and motor control commands. 
To optimize the workflow for experiment design the NRP provides user tools for fully customized setups such as a Robot Designer, Environment Designer, a Virtual Coach for scripted experiments and State Machine for automated control. Communication is based on the widely used Robot Operating System (ROS) \citep{quigley2009ros} that enables modular interchange of information within the NRP, as well as with external processing nodes in soft- and hardware. In this work we add an additional processing node within this modular framework for EMG data processing and control with a Genetic Algorithm.
The NRP is deployed on the EBRAINS computing cluster infrastructure and hereby accessible via web browser. We here employ a local installation of the NRP on a local computer as it is offered open source to a wide user community. Use cases include the evaluation of neural networks based on neurophysiological experiments in embodied closed-loop simulations with musculoskeletal and robotic models. 

The NRP has been used for reconstruction of physical experiments before, in particular a stroke rehabilitation study with mice has been implemented in \citep{mascaro2020experimental}. Here, we focus on the human musculoskeletal system and EMG data of muscle activation that help us better understand biological muscular motion control patterns.

\section{Experimental Setup}

In this section we describe the experimental study that has been realized in this work. We first introduce the participant study of human knee extension and then we describe the simulation architecture that has been implemented to reproduce and evaluate the study data virtually.

\subsection{Human knee extension study}

We conducted a modified knee extension study with human participants as lower limb muscles are
expected to provide the strongest muscle activation feedback in the human body \citep{york2021effect}. Here, we focus on the knee angle alone as the one degree of freedom (DOF) at the rotary joint as all others will be avoided by providing support. One degree of freedom also provided better measurable one dimensional feedback, reducing the effect of extraneous variables.

A cross-sectional, single-blinded control study was designed to compare the muscle activity at four distinct knee angles during isometric contractions. Healthy participants (n = 17, female = 8) within the ages of 18-30 (24.4± 2.57 years) without previous knee joint injury participated in the study. Participants were randomly allocated to different groups, and all measurements were conducted in the motor control laboratories of the center for sports sciences, in the School of Biomedical Sciences, University of Leeds, UK.. 

A physio-clear plastic Goniometer angle ruler was used to measure the knee extension angles. Participants lay supine on the bed with the head, back and leg muscles being fully supported. The setup is adapted from the Fugel-Meyer's knee control test used in stroke rehabilitation \citep{fugel1975}. The knee was held in position using a locking knee brace (DonjoyTM).

\subsubsection{Study Procedure} \label{ssec:exproc}
Subjects were asked to perform an isometric knee extension experiment with maximal voluntary effort. Initially, participants lay suspine on the bed with both legs stretched out in front, the hip, knee and ankle are located at a 0\degree{} to each other. For every recording the participant contracted their muscles at the back of the thigh of the right leg, while the knee is placed at a given steady angle. Participants were asked to activate the Rectus Femoris (RF) voluntarily, to ensure RF was the most active muscle, in line with current literature. We examined four different knee angles: at 0\degree{} (foot extended, so straight at the knee), 20\degree{} (knee is slightly bent with foot pointing away), 60\degree{} (middle of the range of knee flexion),  and 90\degree{} (foot is at a perpendicular to the hip regarding the knee).

The order in which the knee angles were presented for testing was randomized across participants. Overall, participants contracted their muscles at each angle 6 times, with each contraction lasting 5 seconds, with a 3-minute break between each contraction. 

\subsubsection{EMG data recording} \label{ssec:emgrec}
For all the four fixed knee angles with maximal muscle effort EMG data was recorded. For this purpose sensor pads were attached to record from the knee muscles Rectus femoris (RF), Vastus Lateralis (VL), Vastus Medialis (VM), Semitendinosus (Se) and Biceps Femoris (BF). For recording wireless sEMG sensors (Delsys TrignoTM system; at 1.9 kHz, the bandwidth of 20 to 450 Hz) were used. 
Electrode placement and recording sites were selected based on the SENIAM protocol \citep{rainoldi2004}. We used anatomical landmarks to identify the muscles following the SENIAM guidelines \citep{rainoldi2004}. Prior to placement of sEMG electrodes, relevant skin areas were shaved and cleaned with isopropyl alcohol, abraded with preparation gel (Nuprep, NRSign Inc., Canada), \citep{merletti1998repeatability}. The electrode placement was verified in each subject by palpating the muscles and asking participants to perform a muscle contraction. The root-mean-square (RMS) of the evoked muscle activity was then calculated. Activity across muscles was normalized to that in RF, which was voluntarily activated by the participants at each angle in both tasks.

\begin{figure*}
	\centering
	\includegraphics[width=\linewidth]{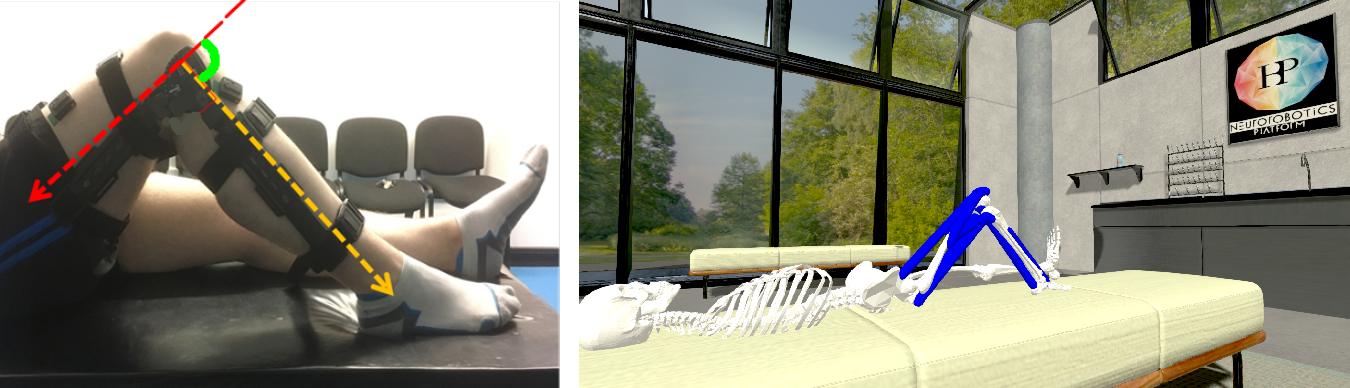}
	\caption[EMG Experiment]{Human knee extension experiment: Participant Study with EMG recordings (Left) and experiment replication in the Neurorobotics Platform (Right).}\label{fig:exp}
\end{figure*}

\subsection{Simulation Architecture}
EMG data analysis reveals insights into correlation between muscle activation and knee angle, but solely looking at the recorded EMG data a conclusion about resulting joint torques cannot be derived. We therefore reconstruct the physical human knee extension study in a simulated environment: We model a virtual room that contains an experiment bed and a musculoskeletal body model as a physics simulation in the NRP. \autoref{fig:exp} shows both an exemplary test setup of the original participant study (left) and its replication with the musculoskeletal simulation as the basis for the experiments described in this paper. 

\begin{figure}
	\centering
	\captionsetup{width=.6\textwidth}
	\includegraphics[width=.5\textwidth]{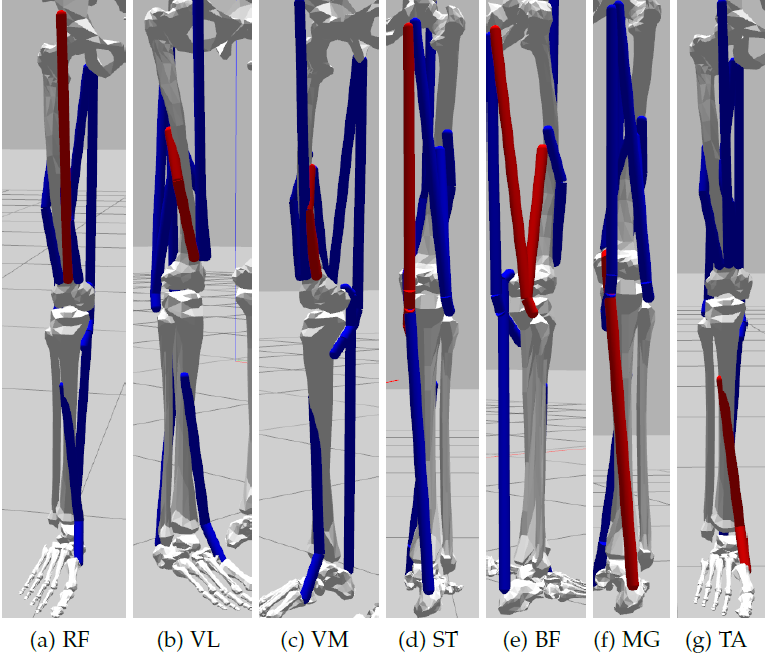}
	\caption[]{Muscles of the simulated lower limbs considered in our experiment: Rectus Femoris, Vastus Lateralis, Vastus Medialis, Semitendinosus, Biceps Femoris, Medial Gastrocnemius, Tibialis Anterior }\label{fig:musclesmodel}
\end{figure}

An available OpenSim model of the lower limb \citep{au2013gait} \citep{Frank.1999} \citep{Delp.1990} is adapted for this study and its implementation in the NRP. Joints are reduced to basic revolute motions and only muscles recorded in the physical study are used. These are visualized in \autoref{fig:musclesmodel}: 
Rectus Femoris (RF), Vastus Lateralis (VL), Vastus Medialis (VM), Semitendinosus (Se), Biceps Femoris (BF) as well  Medial Gastrocnemius (MG) and Tibialis Anterior (TA). All muscles are simulated according to "Thelen2003" \citep{Thelen2003} hill type models that emulate characteristics of elasticity and damping. The torso is fixated to the virtual bed, the right leg can move in a planar space in line with the physical setup.
We implement a ROS control node in Python that controls simulated muscles with an activation value in range [0,1] (0 represents a fully relaxed muscle and 1 expresses maximal activation). Muscle synergies as well as resulting biomechanic behaviors measured in terms of joint angles and torques are recorded for analysis.

\begin{figure*}
	\centering
	\includegraphics[width=\linewidth]{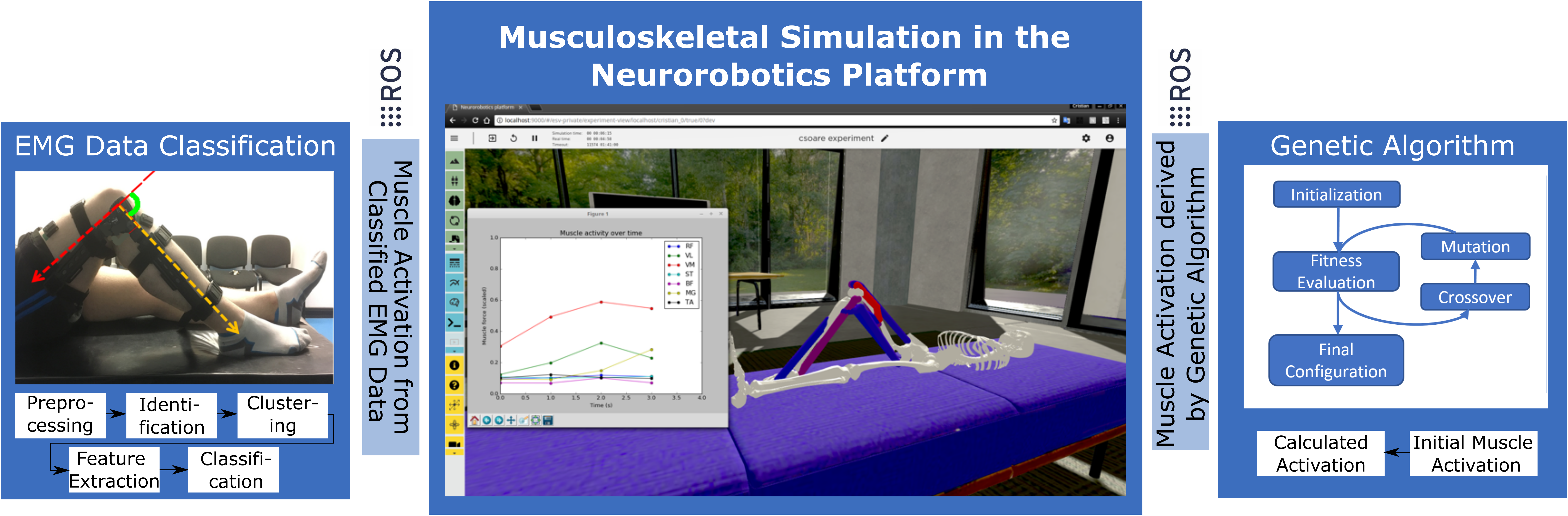}
	\caption[]{Experiment setup in the Neurorobotics Platform with two distinct control approaches: EMG data from the participant study is processed and applied to the simulated model to record resulting knee forces. In a second step optimal synergies are calculated by Genetic Algorithms using the simulation model as test and evaluation bed.}\label{fig:arch}
\end{figure*}

 In this paper we demonstrate a pipeline that consists of two different control and evaluation steps to actuate the musculoskeletal model. First, we calculate normalized muscle activation values from EMG data collected during the participant study. Second, we implement GA to compute an optimal muscle control pattern by active trial on the simulated model. \autoref{fig:arch} visualizes the architecture, that is centered around the virtual experiment replicate, conceptually. In both control steps we interface derived muscle activation data with the physics simulation by means of ROS topics. As a result we inspect and record knee joint angles and torques on the simulated model.

\section{Data Processsing}

Data passes through several steps before the final virtual experiment can be performed. First, the raw EMG data is preprocessed and muscle activations are extracted from the trial recordings. The normalized inputs are then fed into the NRP simulated muscles and the resulting torques are measured. Finally, these torques are taken as reference and a GA is used to generate muscle activation configurations which replicate the measured values.

\subsection{EMG Data Analysis}

\subsubsection{Trial Localisation}
\begin{figure}
	\centering
	\captionsetup{width=.57\textwidth}
	\includegraphics[width=0.57\linewidth]{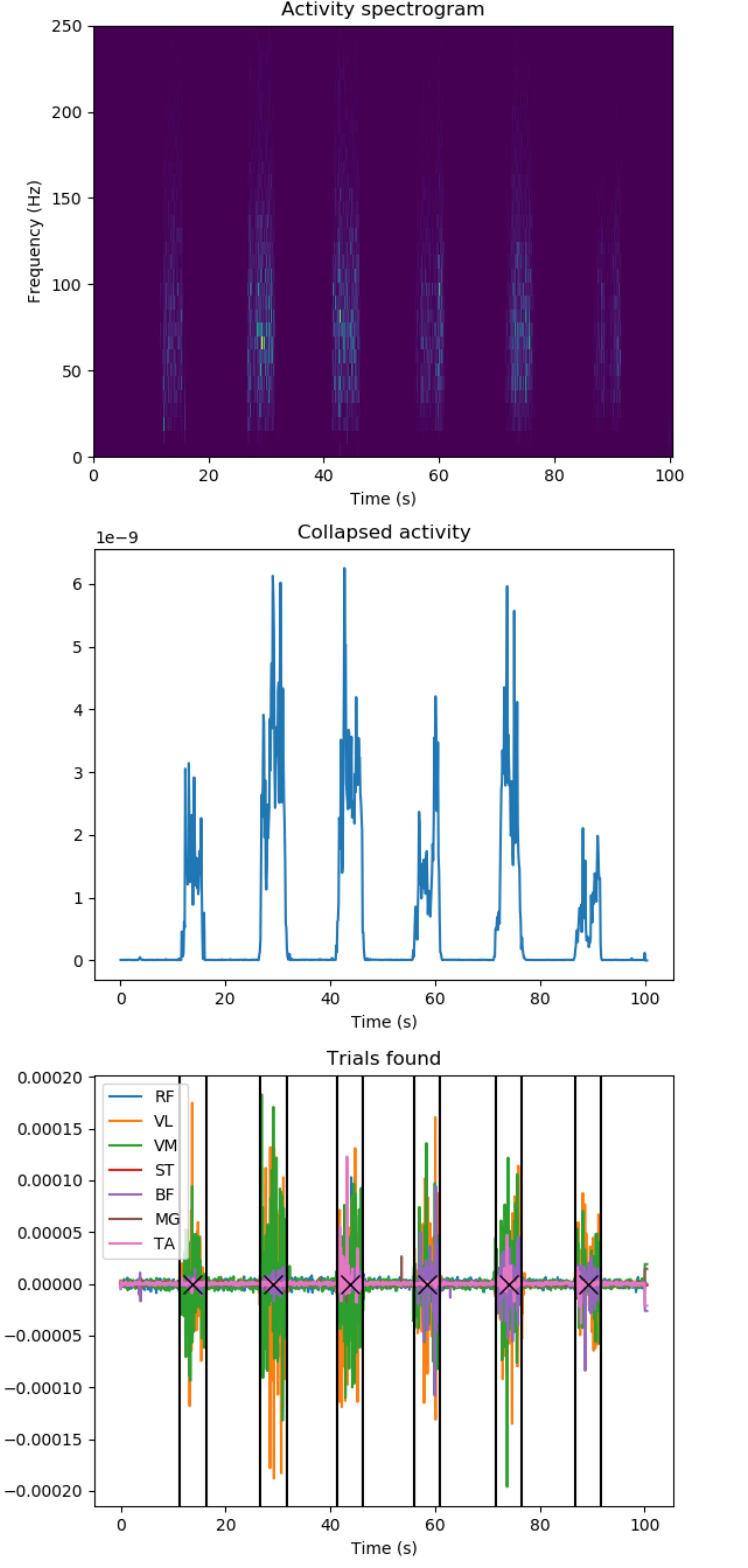}
	\caption[]{EMG data knee extension trial localization: (top) Activity spectrogram of muscle all recordings: Frequency spectograms for all recorded muscles are added up to identify overall muscle activation during a participant trial session. (middle) Summed activity spectrogram: Frequencies are added up at any given time point to spot knee extension trial durations. (bottom) Identified knee extension trials in original recorded EMG data (colors represent individual muscles): A participant executes several trials of knee extension in a row, we apply signal processing to identify trial duration (vertical lines) and centers (crosses). }\label{fig:emg-proc}
\end{figure}

To perform the experiment, the trials have to be isolated and extracted from the original recordings and the EMG signals have to be preprocessed and readied for the simulation. As previously stated in \autoref{ssec:exproc}, the experiments consists of 6 trials, each lasting 5 seconds, with a 3 minute break in between. Signals are first transferred into frequency domain, rendering 7 spectrograms per experiment, corresponding to the 7 muscles of interest. The spectrograms are then summed into a single muscle activity spectrogram \autoref{fig:emg-proc} (top) which, in turn, is collapsed into a 2D plot \autoref{fig:emg-proc} (middle) by further summing the amplitudes over all frequencies at any given time point. Each time the activity plot is greater than a threshold determined by the SNR, an activity event is marked in time domain. Since it is known that each experiment has 6 trials, the K-means algorithm is used to find the trial origin of each activity event by clustering them in 6 clusters. Finally, the median time point of a trial's activity events determine its central time-wise occurrence (\autoref{fig:emg-proc} bottom).

\subsubsection{Signal Preprocessing}

\begin{figure}[H]
	\centering
	\includegraphics[width=0.7\linewidth]{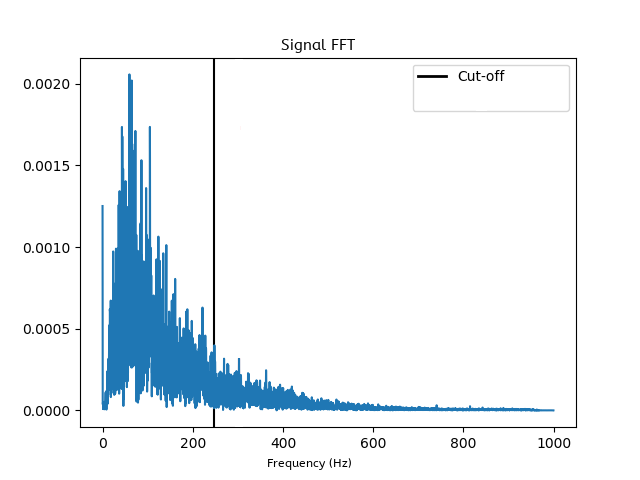}
	\caption[]{Frequency domain plot: The FFT-generated data is averaged over all trials, which shows that the most prominent frequencies lie within the lower end of the spectrum.  All frequencies higher than the computed cut-off point will be ignored in further analysis}\label{fig:signal-fft-cut}
\end{figure}

The data contains noise components which have to be filtered out. Moreover, only a specific bandwidth may offer relevant information for further analysis. By observing the signals in frequency domain, it becomes evident that most high amplitude activity occurs in the lower end of the spectrum, as seen in \autoref{fig:signal-fft-cut}. Therefore, to increase the signal analysis reliability, we need to find a heuristic that automates calculating a cut-off point that is tailored to our data set. The FFT analysis showed that lower frequencies are of higher amplitude and are in fact a minority when compared the rest of the spectrum, a property which has proven useful in analyzing the signal from a statistical angle. The proposed heuristic is as follows: After averaging the FFT values across all study experiments and creating an amplitude-wise histogram of the frequencies, we fit a binomial distribution envelope to said histogram. We then find that the first outlier outside the standard deviation acts as a good amplitude threshold. The highest frequency reaching this threshold will therefore be considered the cut-off point.

\subsubsection{Feature Extraction} \label{sssec:featext}

To be consistent with the study experiment, muscle activation data has to be inferred from the EMG signals. It is known that a muscle unit (MU) exerts a muscle unit activation potential (MUAP) when activated. An increased firing rate of a neuron translates into a higher stimulation of the muscle fibers, resulting in a sustained force (tetanic contraction) in the MU. Furthermore, due to the size principle in MU recruitment (\citep{mendell2005size}), larger neurons, associated with a higher firing threshold are activated with the increase of force, which can be observed as an increase of voltage in the EMG recordings. Taking these into account, our muscle force approximation heuristic equates to $$F = \sum_{f=low}^{high} f \times v(f)$$ 
A weighted sum where each constituent frequency component is multiplied by its amplitude.

\subsection{From Muscle Activation to Simulation Control}

\begin{figure}[H]
	\centering
	\includegraphics[width=0.6\linewidth]{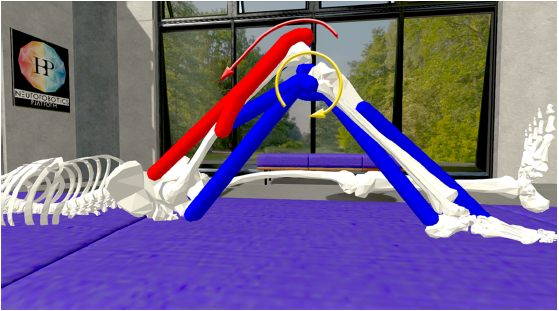}
	\caption[]{Knee torque evaluation on the simulated musculoskeletal model in the Neurorobotics Platform: Muscles are activated (muscle color coding: blue - no activation, red - maximal activation)) by either processed EMG signals or a Genetic Algorithm Muscle generated torque (red arrow). A counter acting knee torque (yellow arrow) is iteratively increased until the goal knee angle is reached.}\label{fig:optorque}
\end{figure}

Once the EMG recordings have been converted to muscle activation scalars, these values are normalized and applied on the NRP leg model muscles. Consequentially, the virtual muscles are contracted and the leg is extended with a resulting knee torque. The NRP does not provide a way to directly measure a muscle generated torque in a given joint. However, it does allow dynamic interaction with the model via its Python API. Therefore, by applying an increasingly stronger virtual torque opposing the muscle generated one (shown in \autoref{fig:optorque}), once the leg reaches its original bent position it can be inferred that the artificial force equals the muscle driven one. The approach of applying torques that are increasing is supported by \citep{beltman2004voluntary}.

\subsection{Genetic Algorithms to Compute Optimal Muscle Activation}

Having extracted the knee torques from the original EMG recordings, using the NRP simulation, we can test whether the same rotational force can be achieved through different synergies. To do this, a GA is used in conjunction with the simulation to evolve an activation pattern that matches the target torque. The GA makes constant use of the simulation to measure the fitness of its population.
In this problem context, each individual has 7 genes representing the activations for the controlled muscles. Its fitness is defined by $cos(|a_{current} - a_{target}|)$, where $a_{current}$ is the knee angle observed in the simulation using the current individual's activations and $a_{target}$ is the target angle of the algorithm.

While this evolutionary approach can find muscle configurations that match the target study experiment torques, it is not guaranteed that any given solution has at least a fully tensed muscle. Each gene within an individual is created randomly with a value between 0 and 1. This means that the configuration does not represent maximal voluntary effort. To mitigate this, the genes of an individual are scaled to the maximal value which becomes 1, while the others increase proportionally.

\section{Results}

\subsection{EMG Data Analysis}

\begin{figure*}
	\centering
	\includegraphics[width=\linewidth]{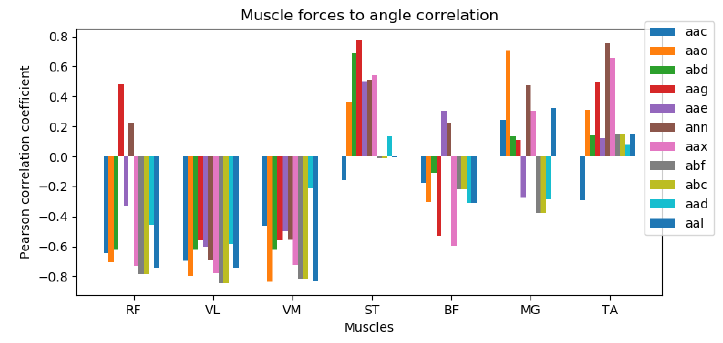}
	\caption[]{Muscle to knee angle Pearson correlations for multiple patients: RF, VL and VM show a strong negative muscle activation correlation with given angle, ST and TA a positive correlation.}\label{fig:muscor}
\end{figure*}

The muscle activation approximation heuristic described in \autoref{sssec:featext} demonstrates results in accordance with physical expectations. As seen in \autoref{fig:muscor}, the quadriceps muscles (Rectus Femoris, Vastus Lateralis, Vastus Medialis) are inversely correlated with the knee angle, while the posterior muscles (Semitendinosus, Biceps Femoris) show weak correlations as they don't take an active part in the leg extension effort.

\subsection{Simulated Knee Extension}

\begin{figure}
	\centering
	\includegraphics[width=0.7\linewidth]{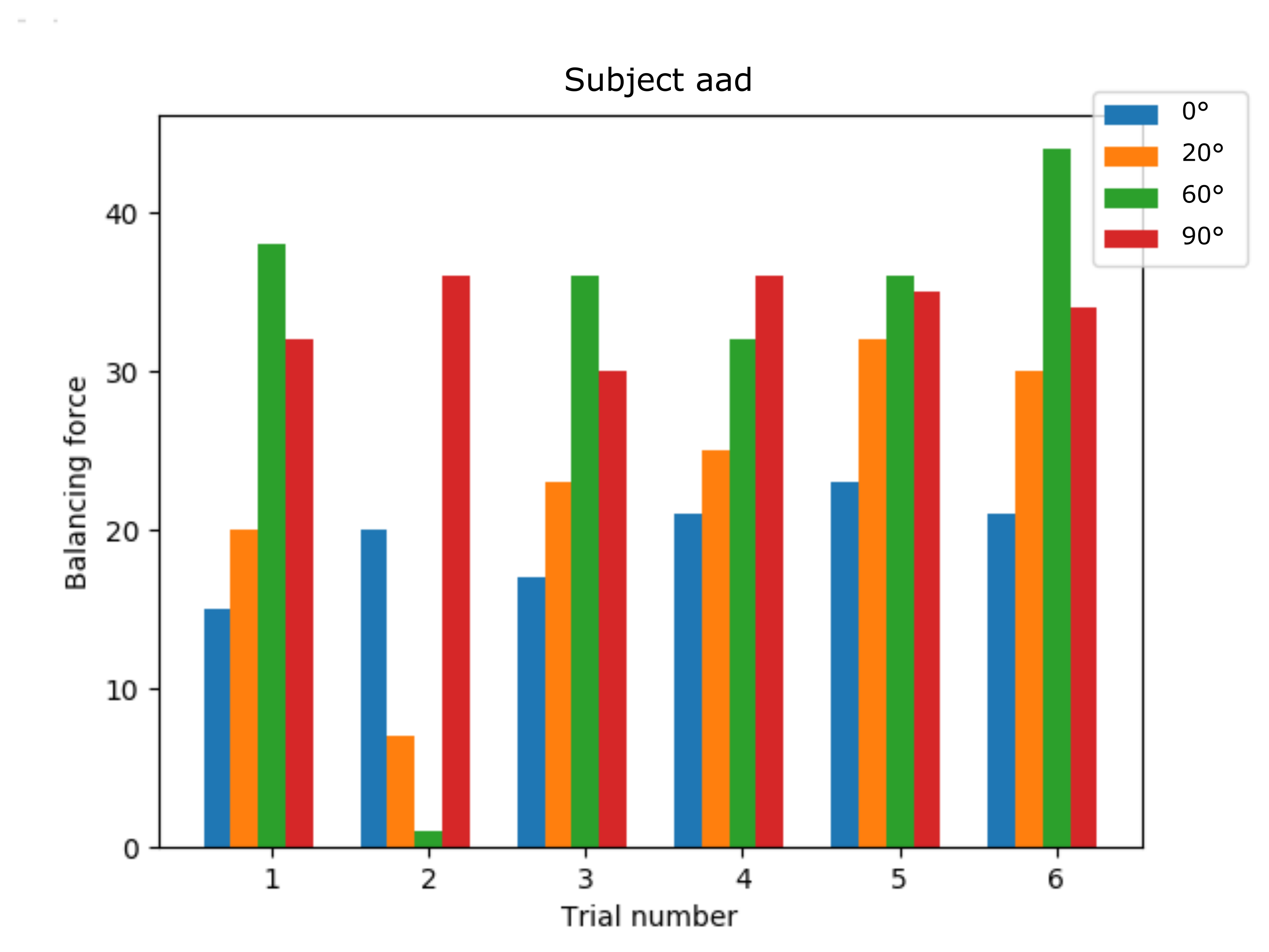}
	\caption[]{Knee torque for subject aad: Generated torques show a common pattern across trials and non optimal trial executions can be identified. }\label{fig:torques}
\end{figure}

The NRP has proven crucial in analysing the resulting knee torque by applying the EMG derived muscle activations. As seen in \autoref{fig:torques}, the simulation is showing under which trials the subject might not have performed the maximum voluntary effort study experiment correctly (i.e. the second 20 and 60 degree trials show below average muscular activation when compared to the rest of the trials).

\subsection{Comparison of EMG derived and GA generated Muscle Activation}

Our experiments have shown that the GA reaches a conclusion fairly rapidly after around 20 generations. However, since the fitness function waits a couple of seconds for the simulation to reach a stable position with the evaluation of new individual, the entire process may take up to tens of minutes.

Comparing natural and evolved muscle activations achieving the same extension torque, configurations are not linearly correlated for any given subject and angle. This is to be expected indeed, as even comparing two experiments belonging to two different subjects, recorded at the same angle does not present any relevant similarity.

\begin{figure}[H]
	\centering
	\includegraphics[width=\linewidth]{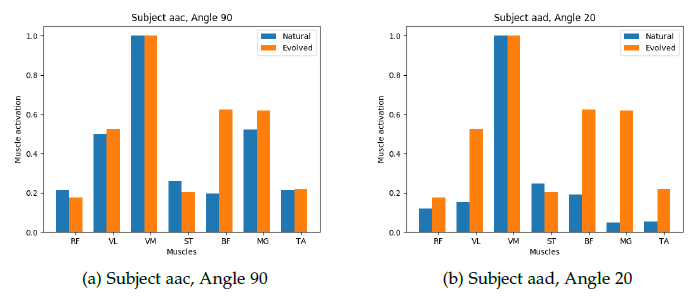}
	\caption[]{Positive comparison of natural vs evolved synergies: The Genetic Algorithm computes synergies that recreate the muscle activation of processed EMG data of the participant study.}\label{fig:evolgood}
\end{figure}

\autoref{fig:evolgood} shows a detailed comparison between the natural and the evolved synergies for two exemplary subjects at at a given angle. The GA found a suitable extension force pattern which closely resembles the original recorded values.

\begin{figure}[H]
	\centering
	\includegraphics[width=\linewidth]{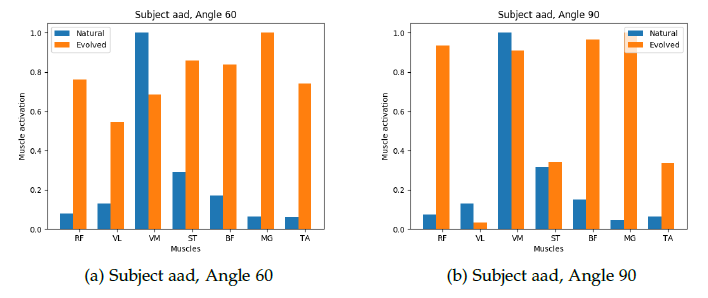}
	\caption[]{Negative comparison of natural vs evolved synergies: The Genetic Algorithm recreates the knee torque of study participants but identifies different synergies to reach the same torque.}\label{fig:evolbad}
\end{figure}

However, \autoref{fig:evolbad} shows that while the algorithm does reach a valid extension effort end state, its results can also end up quite distinct from the original, natural configuration.

The only consistent characteristic in all subjects for both approaches is the overall dominance of the quadriceps extensor over the posterior contracting muscles. Therefore, the conclusion is that the same results, both in pose and in extension torque, can be achieved by different activation patterns.

\section{Conclusion}

In this paper we introduced a simulation framework that uses a virtual musculoskeletal model in the NRP for calculating effective joint torques of recorded surface EMG data. The model skeleton is simulated with rigid body dynamics in Gazebo and extended with hill type muscles of OpenSim. Within our framework we implemented two different control components to actuate the virtual model. First, we analyzed surface EMG data and computed normalized muscle activation to control the simulated model. We could replicate the experimental study virtually and inspected the physical impact of muscle forces and joint torques by applying increasing counteracting torques. Both force values are highly relevant research objectives \citep{zhu2017re} \citep{zhou2011surface} but could only be accessed invasively so far \citep{roberts2008interpreting} \citep{disselhorst2009surface}. Furthermore, our implementation makes use of a physics simulated model in a virtual environment and hereby is more generic than other approaches with e.g. LabView in \citep{zuchruf2021simulation}.
Using this model in a second step we implemented a Genetic Algorithm that computes optimal muscle control configurations that can reproduce observed joint torques. Overall, the framework is implemented using ROS interfaces and design tools provided with the NRP and hereby is highly modular and can easily be applied to a variety of musculoskeletal models and experiment scenarios. 

In this work, we demonstrated the capabilities of our framework and its advantages for EMG evaluation with specific data from a human knee extension study. For this purpose we integrated a human musculoskeletal model of the lower limbs into the NRP and hereby replicated the physical study setup in-silico. In line with the physical constraints applied in the human participant study the simulated leg in the NRP is restricted to motions in a vertical space. However, the simulated model can easily be extended (e.g. in terms of number of muscles, Degrees of Freedom and limb) and hereby opens up the introduced framework for a variety of further studies in the future. Previous work with musculoskeletal models in the NRP focused on rodent motion control paradigms to study stroke rehabilitation in the cortex \citep{mascaro2020experimental}. We investigated neural activation signals to muscles, the lowest level of neural activation in the motion control hierarchy contributing to a better understanding of the descending pathways of motor control.
The data analysis showed that quadriceps muscles are inversely correlated with the knee angle, while the posterior muscles show weak correlations. With the simulated model we demonstrated that even with the same goal knee angles participants applied different muscle configurations to achieve these. 

With our GA approach, we can reproduce this variability of synergies and showcase additional control strategies leading to the same knee torque. The GA suggests solutions that match the experimental  outcomes, but also some unlike what the subjects produced - e.g. in individuals there is maximal activity in primary muscles while the GA also finds solutions with increased activity across multiple muscles. This is maybe because GA is unbiased and not working within the human biomechanical constraints.

Overall, the NRP allowed us for the first time to know and reproduce the effective joint torques and muscle synergies that are generated during a task. The observed variability supports the idea that the muscle interactions are not all hardwired or predefined but fairly flexible even if the endpoint or the torque generated is almost the same across all individuals. The observed muscle recruiting requires a neural control system that is less rigid and responsive to the specific needs, as predicted by \citep{york2021effect} recently, where they show that the proprioceptive inputs are capable of affecting the muscle recruitment pattern depending on the need. Our study shows that the synergies generated are mapped to the specific need too, so the controller is probably modulated by the sensory perception to produce both the knee torque to be generated. This can also explain how to account for the state of the muscle and the level of recruitment of the fibres or state of fatigue that it may be in when conducting a task.

\section*{Conflict of Interest Statement}

The authors declare that the research was conducted in the absence of any commercial or financial relationships that could be construed as a potential conflict of interest.

\section*{Acknowledgments}

This project/research has received funding from the European Union’s Horizon 2020 Framework Programme for Research and Innovation under the Specific Grant Agreement No. 785907 (Human Brain Project SGA2) and No. 945539 (Human Brain Project SGA3). PS was funded by the Royal Thai Government Scholarships.

\section*{Study Details}
This study of knee flexion was conducted with healthy participants (n = 17, female =8) within the age of 18-30 (24.4± 2.57 years). All participants were healthy and free on any injuries, fully informed about the tests, and all gave their written, informed consent before participation. The study was conducted according to the declaration of Helsinki and was approved by the Local Ethics Committee of the University of Leeds (reference number BIOSCI 16-004). Participants were excluded if they d 1) recently done exercise within 48 hours prior to testing 2) Knee stiffness, self-reported pain 3) Muscle or knee joint injury 4) Used recreational or performance-enhancing drugs 5) Ingested alcohol in the previous 24 hours 6) Unable to provide informed consent. The data were recorded directly onto a PC running Spike 2.8. version 10 (Cambridge Electronic Designed limited, (CED)) software using the Delsys talker (CED) for later offline processing and analysis.

\section*{Author Contribution}
BF conceptualized the experiment design of EMG data evaluation in a virtual simulation environment exploiting the NRP simulation in discussion with MdK. PS conducted the participant study and collected the EMG data under supervision of SC and SA. Human experiments were designed by PS and SC. AK supervised the NRP development and provided input for the given use case. CS and BF analyzed the data and implemented the experiment in discussion with SC and MdK. BF wrote the paper with text input from MdK and SC for the introduction and EMG data collection, as well as CS for the analysis and implementation.

\bibliography{references}
\end{document}